# Viscosity and dynamic surface tension measurement: A guideline for appropriate measurement.[†]

Vivek Kumar[a], JSM Quintero[a], Aleksey Baldygin[a], Paul Molina[a], Thomas Willers[b], Prashant R. Waghmare[a,c]

[a]interfacial Science and Surface Engineering Lab (iSSELab), Department of Mechanical Engineering, University of Alberta, Edmonton, T6G 1H9, AB, Canada
[b]KRÜSS GmbH, Borsteler Chausse 85, Hamburg, 22453, Germany
[c]Corresponding author, email:prashantwaghmare@cunet.carleton.ca,

**Abstract**

Dynamic surface tension measurements play a critical role in interfacial activities for liquids with varying viscosities. Understanding the rate at which the interface attains the equilibrium, for surface tension measurements, after the formation of a new interface is of significant interest. Although surface tension is independent of viscosity, the time required for a new surface to form (equilibrium/relaxation time) is influenced by viscosity. The measured surface tension value is accurate only once these viscous effects have subsided. Therefore, the relaxation time represents the minimum surface age value achievable during the measurement process. We experimentally established the minimum surface age needed to measure the dynamic and static surface tension of a fluid with a specific viscosity using three widely used methods: the Pendant drop method, the Wilhelmy plate method, and the Bubble pressure method. We propose a guideline with a phase plot that helps to choose the most suitable method and the youngest achievable surface age for an accurate measurement, independent of viscous effects. This guideline enables users in diverse applications such as 3D printed clothing, spray paint, coating, etc., to accurately measure dynamic and static surface tension without being influenced by viscosity effects.

*Keywords:* Dynamic surface tension (DSFT), Surface tension (SFT), Viscous effect, Equilibrium/relaxation time, Surface age, Hydrodynamic forces

## 1. Introduction

Dynamic surface tension (DSFT) is important in diverse applications such as coating manufacturing, sports apparel, 3D printing, cannabis extraction and processing of byproducts, ink-jet and resin-based lacquers [1, 2, 3, 4, 5, 6, 7, 8, 9, 10, 11]. In many of these application processes, alteration of DSFT becomes essential for the desired outcome [12]. The accurate determination of DSFT at the youngest surface is crucial for various futuristic applications such as microfluidics and Lab-on-a-chip devices, biomedical applications, advanced materials and coatings, printed electronics, and environmental remediation [13, 14, 15, 16, 17, 18, 19, 20]. The measurement of the surface tension as function of time passed after the generation of a new surface is called dynamic surface tension. The surface tension is measured at different so-called surface age, *i.e.*, time passed after the surface generation. For pure liquids the surface tension does not exhibit any dynamic behavior and does not depend on the surface age. For liquids containing surfactants the surface tension depends on the diffusion properties and concentration of these surface-active molecules [21, 22, 23, 24, 25].

In the case of a surfactant sample, when an interface is formed, the surfactants from the bulk travel to the interface and occupy the interface. As a result, the DSFT decreases with surface age and ultimately plateaus [26, 27]. Therefore, the effect of surfactant properties and concentration on DSFT is extensively studied [21]. For DSFT measurements, a new surface has to be created in order to measure the surface tension as a function of surface age. However, surface tension is only defined for surfaces in equilibrium and the time for such equilibration after forming of a new surface (relaxation time) depends on the viscosity, primarily for pure, chemically non-reactive liquids. As a result, the measured value is only valid when the interface formation has no dynamic effect due to the viscous nature of the liquid, and thus this relaxation time limits the youngest possible SFA (surface age) accessible during the measurement. Any surface tension reading at a surface age smaller than the relaxation time is not correct. The relaxation time is the limit value for the smallest possible SFA during the measurement. As the way the new surfaces are created is different for different measurement methods, the smallest accessible SFA is not only viscosity but also method dependent.

Numerous studies have encountered similar challenges regarding the influence of viscosity on dynamic surface tension, particularly in relation to the viscosity's effect on dynamic surface tension using the Pendant Drop (PD) [28, 29, 30, 31]. The PD method relies on the deformation of a hanging droplet due to the balance between surface and body forces, expressed through the shape factor [32]. However, during the initial surface generation stage, hydrodynamic forces, generated while plunging the liquid from the syringe to the needle tip, crucially impact droplet deformation, potentially leading to a misleading shape factor. In a preliminary study, Karbaschi et al. (2012) [33] examined hydrodynamic relaxation in PD-based dynamic surface tension measurements, emphasizing the roles of flow rate and droplet volume. However, their analysis was restricted to low-viscosity systems (e.g., water/air, water/hexane) and did not address the broader viscosity dependence. The full extent of this phenomenon's impact remains unclear [34, 35, 36].
Regarding the Wilhelmy Plate (WP) method, the influence of viscosity on dynamic contact angle measurements and surface tension is well-established [37, 38, 39]. Data for the WP method were directly extracted from our previous work, as cited by Rahman et al. (2019) [37].

In the case of Bubble pressure (BP) method, hydrodynamic forces significantly influence measurements, especially at small surface ages [40]. Some studies have provided corrections for this effect, but they are often specific to certain liquids and lack details on viscosity variation. The foundational work by Miller and co-workers on the BP technique, widely referred to as Miller's Maximum Bubble Pressure Tensiometry (MBP) method, primarily utilized complex, non-commercial setups that required correction for minimum measurable time—referred to as the *dead time* ($t_d$) [12, 25, 41]. Defined as the interval between the maximum and minimum pressure during bubble formation, $t_d$ accounts for instrumentation-related delays in capturing accurate surface tension values. While $t_d$ may appear conceptually similar to the relaxation time, it is fundamentally different: $t_d$ is setup-dependent and does not capture the dissipation of transient hydrodynamic effects. Therefore, for accurate MBP measurements, the minimum allowable time between successive readings should account for both the instrument-specific dead time and the fluid-specific hydrodynamic relaxation time. Similar to these studies, an early experimental study presented correction factors for the surface tension of a water-glycerin mixture over time [22]. Fainerman et al. (1993 & 2004) [22, 42] investigated the maximum bubble pressure of water-glycerin mixtures and found that dynamic surface tension measurements were influenced by viscosity. However, their work provides a correction factor for water-glycerin mixtures without stating the equilibrium time. Moreover, implementing the findings of such studies across broader scenarios is challenging. Despite extensive research in both the PD and BP methods, the full extent of viscosity's influence and proposed guidelines applicable to a wider range of viscous fluids are yet to be fully established. To account for these factors and correct the DSFT - measured values, studies have been constrained to specific experimental configurations and fluid types, with only a few addressing BP [22, 42, 41, 43, 44, 45], fewer still addressing WP [37, 46, 47, 38, 39], and none addressing PD [28, 48]. Importantly, no study has combined all three measurements to assess the implications of viscous effects on the measurement techniques, nor proposed limiting regimes for each technique based on different operating parameters, using fluids with viscosities ranging from 1 to 1000 $m$Pa.s.

In contrast to the individual work presented for a specific technique, our intent is not to correct for viscosity effects but determine the range of surface age in which no correction factor is needed as viscosity does not affect the measurement anymore. We performed dynamic surface tension measurements of pure liquids, as for those the surface tension theoretically should not depend on surface age. The use of pure, single-component, not reactive with surrounding medium liquids to isolate hydrodynamic relaxation effects without interference from surfactant adsorption or diffusion. In surfactant-based systems, viscosity directly influences the adsorption kinetics via diffusion coefficients, introducing additional timescale that can obscure viscosity-specific hydrodynamic behavior. By comparing the obtained measured (and partly time dependent) data with the expected for a surface age independent behavior we can clearly pinpoint to the surface ages at which the surface tension measurement is falsified by viscosity effects. We specifically focused on commonly used techniques such as the PD, WP, and BP methods. It is worth mentioning that MBP technique developed by Miller is rarely used by BP instrument providers in the market. To conduct our analysis, we utilized viscous liquids of eight different viscosities and incorporated the variation of process parameters such as two flow rates in PD and five capillary diameters for BP. Our study involved measuring dynamic surface tension (DSFT) variation with surface age and determining the critical surface age required to achieve equilibrium surface tension. Furthermore, the use of pure, single-component liquids under carefully controlled conditions allows for the precise isolation of hydrodynamic effects. Furthermore, we compared our BP and PD results with our previous WP results [37]. This comparison of different methods identified the most suitable measurement method and the minimum surface age for accurate DSFT for varied viscosity liquids.

## 2. Background: Measurement principles

As mentioned earlier, we aimed to investigate the role of hydrodynamic effects in three key measurement techniques, namely, BP, WP, and PD. The PD technique measures surface or interfacial tension by balancing gravity with surface tension. Based on the radii of curvature, length scale from the image, and liquid density, the Young-Laplace equation is solved to quantify the static surface tension [32, 49, 50]. In the case of the WP method, the interface is pierced or formed along the plate or ring and how viscosity affects the data is discussed in Ref [47] and is experimentally elaborated in Ref [37]. The BP measures pressure inside the bubble while it is being formed, and at maximum pressure with minimum curvature, the surface tension is determined using the Young-Laplace equation. It is evident these three techniques have different working principles; PD is based on image analysis, whereas WP and BP are force and pressure measurements, respectively. PD and BP facilitate the measurement of the dynamic surface tension as here the respective surface ages are well defined.

In PD a new droplet is generated, usually as fast as possible, at the tip of a needle. Here we define the time at which the final drop volume is dosed into the pendant droplet as zero for the surface age. From that time on the dynamic surface tension is measured from the drop shape (having a fixed volume) at different surface ages. As the generation of a new drop usually takes about 1 s, this time reflects the accuracy of the surface age determination in pendant drop measurements. In BP a continuous gas flow constantly at constant pressure creates new bubbles at a capillary immersed into the liquid. The time it takes to create a new hemispherical bubble at the capillary defines the surface age of the measurement. Different surface ages can be assessed by adjusting the gas flow rate. Here the accuracy of the surface age determination is usually only limited by the temporal evolution of the pressure reading inside of the bubble and better than 1 ms.

## 3. Materials and methods

To achieve a variation in viscosity spanning three orders of magnitude, we meticulously selected set of test liquids, comprising distilled de-ionized water (Milli-Q A10, Millipore), general-purpose viscosity standards, namely silicone oil (D10, N35, S60, D500, and D1000, manufactured by Paragon Scientific Ltd.), along with silicone oil AP100 (C985N24, Sigma Aldrich) and paraffin oil (18512, Sigma Aldrich). The viscosity of each liquid was assessed at a standard temperature of 20°C utilizing a rheometer (Rheolab QC, Anton Paar) with a double gap concentric cylinder measuring system (DG42, Anton Paar). All relevant thermophysical properties are provided in Table 1. The viscosity standard liquids, primarily composed of mineral oils, possess a certified chemical purity exceeding 99.8%. To rigorously validate their suitability for interfacial characterization, we performed independent in-house measurements of surface tension and viscosity across two production batches. The resulting dynamic behavior remained consistent within experimental uncertainty. All data presented in this study were acquired using a single batch per fluid to eliminate any confounding influence of batch-to-batch variability. The same silicone and paraffin oil standards, sourced from the same manufacturers, have been reliably employed in previous studies on dynamic surface tension and interfacial phenomena [37, 51, 52, 38, 53], further substantiating their relevance and robustness for the present study. The deionized water used showed consistent equilibrium SFT (71.8 ± 1 mN/m) across batches and labs, confirming that trace impurities had no significant impact.

| Liquid | $\rho$(kg/m$^3$) | $\mu$($m$Pa.s) | $\sigma$($m$N/m) |
|---|---|---|---|
| Water | 1000 | 0.89±0.1 | 71.8±1 |
| D10 | 846.2 | 10.32±0.2 | 28.9±0.3 |
| N35 | 853.6 | 59.19±0.1 | 30.8±0.1 |
| Silicone oil AP 100 | 1006 | 100±0.7 | 20±0.1 |
| S60 | 857 | 100.6±0.4 | 31.2±0.1 |
| Paraffin oil | 827 | 140±0.7 | 26±0.2 |
| D500 | 869.9 | 541.4±1.0 | 31.9±0.2 |
| D1000 | 870.7 | 1138±1.3 | 32.2±0.1 |

Table 1: Thermophysical properties: density ($\rho$), viscosity ($\mu$) and equilibrium surface tension ($\sigma$) of sample liquids measured at room temperature (20°C). The surface tension provided is the equilibrium mean and deviation of from all three methods.

For the pendant drop method, a commercial goniometer (DSA30E, KRUSS Scientific Instruments Inc.) capable of measuring surface and interfacial tension is utilized. The measurement range for this device is from 0.01 to 2000 $m$N/m with a resolution of 0.01 $m$N/m. The software (ADVANCE, K R Ü SS Scientific Instruments Inc.) extracts the curvature of the drop and employs the Young-Laplace to determine the surface tension. The dispensing flow rate used to generate the drop is set to 15 and 25 $\mu$L/s, and the needle diameters are of 1.8 mm (*OD-I*) and 2.0 mm (*OD-II*), commonly used needle size for pendant drop measurements. Needle OD (outer diameter) was verified using a digital micrometre (906.050, Schut). The volume for the measurement using the pendant drop was initially set to be as large as possible, just below the point where the drop detaches from the needle. For correctness of measurements, the Bond number ($Bo = \frac{\rho g R_0^2}{\sigma}$) is verified to not touch limit criteria of PD method (i) $Bo \gg 1$ and (ii) $Bo \simeq 0$ [48, 54].

In the Wilhelmy plate method, a commercial force tensiometer (K100, KRÜSS Scientific Instruments Inc.) with a measurement range from 1 to 2000 $m$N/m and a resolution of 0.001 $m$N/m is used. A platinum plate (PL01, KRUSS Scientific Instruments Inc.) is meticulously cleaned with acetone and toluene and then heated with a Bunsen propane burner to ensure consistency and cleanliness. In WP measurements, the plate is often recommended to be immersed by 2 $mm$ into the fluid and then retracted to 0 $mm$ immersion depths. This should further ensure full wetting of the plate by the fluid. In our study, we skipped this immersion step to observe the purely surface tension and viscosity driven flow of the liquid lamella to its equilibrium shape [37]. DSFT using WP can be influenced by wetting-kinetics effects (contact-angle evolution and surface microtexture). This was addressed by: (i) using the instrument's standard platinized (roughened), high surface energy platinum plate (roughness not varied) and flame-cleaning it between runs; and (ii) performing measurements at zero immersion depth, ensuring near-complete wetting ($\theta \approx 0°$) and minimizing protocol/wetting-induced artifacts; under these conditions, the relaxation time is only viscosity-dependent [37].

A commercial bubble pressure tensiometer (BP100, KRÜSS Scientific Instruments Inc.) with precise control over the surface age from 5 to 200, 000 $m$s is used. This allows the surface tension measurement from 10 to 100 $m$N/m with a resolution of 0.01 $m$N/m. The submersion depth up to which the capillary is immersed inside the liquid to a depth of 10 $mm$, which ensures the consistency of the results. This consistency maintained the same magnitude of hydrostatic pressure due to the liquid column above the capillary tip. To examine the impact of capillary radius on the dynamic surface tension results, five different capillaries are used, including glass capillaries of various diameters, such as *SH*2030, *SH*2031 (smallest), *SH*2037, *SH*2040 (largest), and *SH*2510, with diameters of 0.225, 0.205, 0.302, 0.433, and 0.353 mm, respectively. To ensure measurement consistency, all glass capillaries were hydrophobically modified (silanized), following established protocols [55, 56]. Proper wetting and stable tip geometry are essential for clean bubble detachment and artifact-free MBP measurements. Capillaries were sourced from a single batch and regularly monitored to eliminate surface property variations and ensure reproducibility across all tests. To avoid contamination, the capillaries are cleaned with hot running water and isopropyl alcohol (IPA) before and after each measurement. For error analy-sis, each result is produced in triplicate for each test method and experimental configuration.

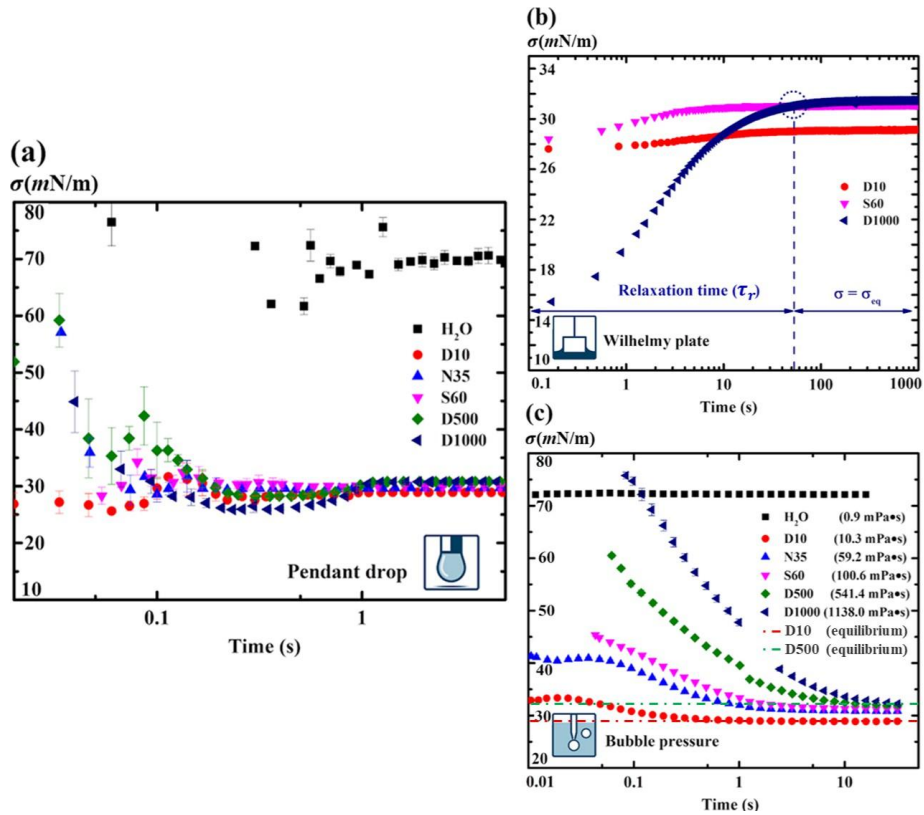

Figure 1: Surface tension with surface age for the (a) pendant drop, (b) Wilhelmy plate, and (c) bubble pressure method.

## 4. Results and discussions

*4.1. Influence of viscosity on measured surface tension values*

In this study, the relaxation time($\tau_r$) is defined as the time required for a process to attain a state of equilibrium, indicating the minimum surface age necessary for stable or equilibrated measurements. Mathematically, the relaxation time ($\tau_r$) is determined as the duration required for measurements to stabilize at an equilibrium value while maintaining an error deviation of less than ±1%. Figure 1 shows the transient variation in the surface tension readings with three different methods: (a) pendant drop, (b) Wilhelmy plate, (c) bubble pressure for eight different viscosities. The error shown in Figure 1 represents the deviation arising from five measurements.

For pendant drop (PD) measurements (refer to Fig. 1), the highest surface tension at the youngest measured surface age was recorded for the most viscous fluid, D1000, at approximately 60 mN/m—nearly twice the equilibrium value of 32.7 mN/m. In contrast, for the lower-viscosity fluid D10, the maximum measured surface tension was about 35 mN/m, deviating by only 20% from its equilibrium value. These results indicate that higher-viscosity fluids exhibit larger transient deviations in surface tension due to stronger viscous effects at early surface ages. Dynamic surface tension measurements using the PD method rely on a delicate balance between surface and body forces. At short surface ages, hydrodynamic (viscous) forces act against droplet deformation and enhance resistance to interface evolution. These forces lead to a reduced shape factor ($\beta$), artificially increasing the calculated surface tension, as given by the relation $\sigma = \Delta \rho g R_0^2 / \beta$ [32, 48, 54]. As viscous effects dissipate over time, the droplet shape stabilizes and the measurement reaches equilibrium. Thus, fluids with higher viscosity exhibit greater resistance to deformation and longer stabilization times.

Surprisingly, the measured relaxation time ($\tau_r$) in PD experiments shows limited sensitivity to fluid viscosity across the experimental range. For instance, D1000 and D10 exhibit comparable relaxation times of approximately 1.0 s and 0.9 s, respectively, despite their order-of-magnitude difference in viscosity. Even water, with a viscosity several orders of magnitude lower than D1000, displays a similar relaxation time. This observation underscores the weak dependence of $\tau_r$ on viscosity in the 0.1–1000 mPa·s range under the current experimental conditions. For low-viscosity fluids such as water and D10, inertial forces dominate over viscous dissipation. Since inertial forces scale with droplet volume and surface forces scale with surface area, such systems exhibit oscillations in droplet shape. These oscillations cause fluctuations in the shape factor ($\beta$), leading to transient variations in dynamic surface tension. For instance, the $\tau_r$ and associated error bars for water are notably higher than D10. This discrepancy arises from water's high surface electric potential [57] and the pronounced droplet oscillations during PD measurements, driven by its exceptionally low viscosity, as evident in Figure 1 and 2. As viscosity increases, inertial oscillations dampen more rapidly due to enhanced viscous dissipation, initially reducing $\tau_r$. Beyond a certain viscosity threshold, however, viscous effects dominate, leading to an increase in $\tau_r$. This complex interplay between inertial and viscous forces explains the observed plateau in relaxation time and suggests that for much higher viscosities (> 1000 mPa·s), $\tau_r$ may increase significantly—highlighting a potential avenue for future exploration.

Figure 1(b) presents the results of the Wilhelmy plate method [37]. In the pendant drop method, the highest surface tension was at the beginning of the measurement, whereas with the Wilhelmy plate, the observation is the opposite. Despite similar thermophysical properties, D1000 and D10 result in quite different minimum measured surface tensions of 14 *m*N/m and 28 *m*N/m and their relaxation times ($\tau_r$) are 84 s and 166 s, respectively.

With the Wilhelmy plate method, as soon as the very high surface energy plate touches the liquid interface, the liquid begins to wet the plate and form a lamella exhibiting a zero-degree contact angle with the plate. In the final equilibrium shape when the contact angle is 0° the surface tension is acting vertical and thus the force measured by the balance is maximum and reflecting the true surface tension. However, during the time of wetting the contact angle of the liquid lamella on the plate is > 0° and the force measured by the balance in the vertical direction is only a fraction of the true surface tension. The wetting and spreading of more viscous liquids is slower, resulting in a longer time for equilibration. Therefore, D1000 takes at least three to four times longer than D10 to attain equilibrium.

For bubble pressure, as illustrated in Figure 1 (c), similar to the pendant drop method, the DSFT measurement during the early surface age period is observed to be considerably higher than the equilibrium value. To illustrate this difference Figure 1 (c) shows also the expected theoretical dynamic surface tension values for D10 and D500 as dashed line. For D10 and D1000, the maximum measured values differ significantly, 35 $m$N/m and 75 $m$N/m, respectively. And the relaxation times are distinct, 0.9 s and 30 s, respectively. This deviation can be attributed to the underlying principle of the measurement technique. As air passes through the capillary into the liquid bath to form a bubble, the process involves the displacement of liquid hindered by viscous shear forces. Consequently, the pressure required to initiate this displacement needs to be higher (i.e., $\Delta P_{\text{actual}} > \frac{\sigma}{r_c}$), where $r_c$ represents the capillary radius, resulting in an elevated surface tension measurement. Furthermore, in the case of highly viscous liquids and measurements at small surface ages, the contribution of these viscous forces becomes more pronounced. Once the surface age surpasses the dissipation of viscous forces, the results reflect accurate measurements.

The results depicted in Figure 1 demonstrate that regardless of the measurement method employed, the measurement deviation is more significant for more viscous fluids. Furthermore, it is noteworthy to mention that the time required to achieve equilibrium varies among the three methods and by order of magnitudes. Specifically, for viscous liquids, the pendant drop method requires the least amount of time ($10^0$), followed by the BP method ($10^1$), and the Wilhelmy plate method necessitates the longest duration ($10^2$). Thus, it is crucial to understand the quantitative relationship between equilibrium time and viscosity and to establish guidelines for selecting an appropriate measurement method based on the fluid's viscosity.

*4.2. Dependence of Relaxation Time ($\tau_r$) on Viscous Time ($\tau_v$) and Inertial ($\tau_i$) Scales*

It is imperative to understand the correct measurement time in relation to the thermophysical properties of the test liquid. Therefore, conducting a separate analysis of the functional dependency of relaxation time on either viscous and/or inertial forces is crucial. Additionally, to determine the role of density and surface tension, the relationship between relaxation time and density (i.e., inertial time scale) is also verified through experimental data presented in Figure 1 of the Supplementary section. To estimate the impact of inertial forces, their relationship to the inertial time scale, as outlined in Eq. 1, can be assessed. Notably, the time required for a specific liquid to attain a steady state is an independent function of its density and surface tension. Hence, the $\tau_r$ is found independent of $\tau_i$ (refer Figure 1, comprehensive discussion in Supplementary section).

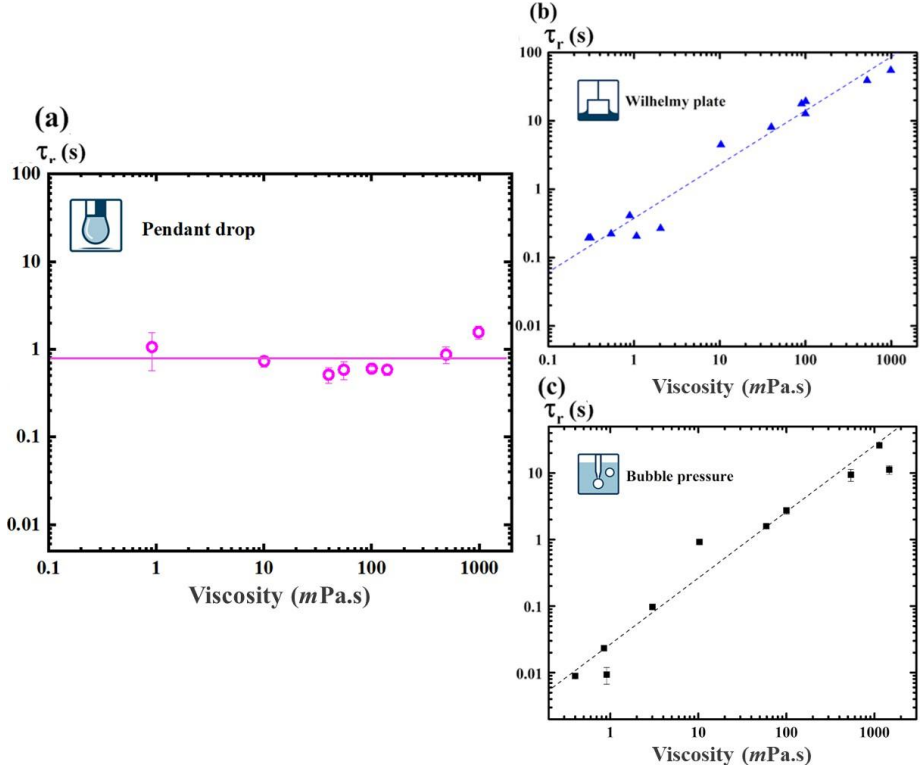

Figure 2: Variation of $\tau_r$ with viscosity ($\mu$) for (a) Pendant drop (b) Wilhelmy plate (c) Bubble pressure methods

$$\tau_i = \sqrt{\frac{\rho r_c^3}{\sigma}}, \quad \text{and} \quad \tau_v = \frac{\mu r_c}{\sigma} \qquad (1)$$

Similar to effect of inertial forces, the effect of viscous forces is identified using the viscous time scale consideration. This scale, represented by Eq. 1, is proportionate to the dynamic viscosity and allows us to examine the variation that occurs with changes in viscosity. Results from experimental measurements of three methods presented in Figure 2 and Figure 3 of the Supplementary section illustrate that the relaxation time of viscous fluids significantly changes with viscous time scales. Furthermore, an increase in relaxation times is observable with an increase in viscous time scale. These findings in Supplementary section indicate that relaxation time depends only on the viscous time scale but remains independent of the inertial time scale.

Establishing a relationship between relaxation time and viscosity is imperative to generate a comparison plot.

*4.3. Empirical correlations between Relaxation Time ($\tau_r$) and Viscosity ($\mu$)*

This section primarily focuses on establishing an empirical correlation between relaxation time ($\tau_r$) and viscosity ($\mu$) within the viscosity range of 1 to 1000 $m$Pa.s. In the low viscosity regime ($\mu$ < 1 $m$Pa.s), inertia assumes a dominant role, leading to oscillations in the liquid-air interface. Consequently, the measurement of surface tension becomes highly sensitive to these interface oscillations, causing fluctuating values in the measured surface tension. This, in turn, presents challenges in accurately determining the exact value of $\tau_r$ using all three measurement methods. Therefore, the findings derived here are not applicable to the low viscosity regime ($\mu$ < 1 $m$Pa.s), providing an intriguing avenue for future studies.

For viscosities greater than 1 $m$Pa.s, the relaxation time is calculated using mean of all measurements for each method. The maximum calculated deviation from mean is always less than 5% irrespective of measurement parameters and method as shown in Figure 2. Using the results, a curve is plotted against $\tau_r$ and $\mu$ for each method. The $\tau_r$ variation with $\mu$ is observed to follow power law due to exponential nature of viscous or hydrodynamic forces. The variation of these hydrodynamics forces are studied by Fainerman et al. (1993), Kaully et al. (2007) and Freer et al. (2005) [22, 58, 59].

When the variation plot is depicted on a logarithmic scale, a distinct linear relationship between relaxation time and viscosity becomes evident for all three methods, as illustrated in Figure 2. The figure demonstrates well-fitted linear variations of $\tau_r$ with $\mu$ each displaying different positive slopes and intercepts. The linear relationships are expressed in the form of general straight lines as shown in Eq. 2, where $y$ represents $\log_{10}(\tau_r)$ and $x$ represents $\log_{10}(\mu)$. The constants $m$ and $b$ represent the slope and y-intercept of the lines, respectively.

$$y = mx + b \qquad (2)$$

The values of $m$ and $b$ are empirically determined from the fitted curves for PD with two flow rates, WP and BP with five different capillaries, alongside the coefficient of determination ($R^2$), and are provided in Table 2. The minimum $R^2$ is 0.91 for these fitted curves, indicating a very strong trend in the experimental data.

| Methods | $m$ (slope) | $b$ (intercept) | $R^2$ |
|---|---|---|---|
| PD (*OD-I*, 15 μL/s) | 0 | −0.06 | 0.91 |
| PD (*OD-I*, 25 μL/s) | 0 | −0.064 | 0.94 |
| PD (*OD-II*, 15 μL/s) | 0 | −0.058 | 0.96 |
| WP | 0.79 | −0.42 | 0.96 |
| BP (SH2031) | 1.01 | −1.62 | 0.94 |
| BP (SH2040) | 0.87 | −1.65 | 0.97 |

Table 2: Variation in rate of change of relaxation time ($m$) with viscosity and intercept ($b$). OD-I capillary dia=1.8 mm and OD-II capillary dia=2 mm. The PD and BP measurements are also performed for two flow rates and two different capillaries.

The parameter $m$ signifies the rate of change of $\log_{10}(\tau_r)$ with $\log_{10}(\mu)$. The positive values of $m$ observed for all methods and measurement parameters indicate a positive correlation, showing that as viscosity ($\mu$) increases, the relaxation time ($\tau_r$) also increases. A higher positive value of $m$ suggests a heightened sensitivity of $\tau_r$ to changes in viscosity. The y-intercept $b$ represents the value of $\log_{10}(\tau_r)$ when $\log_{10}(\mu)$ = 0, indicating the relaxation time of a fluid with a viscosity of 1 $m$Pa.s.

In the pendant drop method, increasing the flow rate from 15 μL/s to 25 μL/s continues to demonstrate independence from viscosity ($m \approx 0$), though it results in a reduction in the parameter $b$. Slightly lower relaxation times are observed at 15 $\mu$L/s compared to 25 $\mu$L/s. The higher flow rate imparts greater momentum to the liquid, leading to increased oscillations, particularly noticeable during measurements of low-viscosity liquids such as water and D10. Consequently, this higher flow rate prolongs the stabilization time due to reduced viscous dissipation. Additionally, increasing the outer diameter (OD) of the capillary also increases the parameter $b$, again due to enhanced oscillations (refer Table 2). The rise in oscillations occurs because inertial forces scale with the drop volume, whereas surface forces scale with the drop perimeter. However, these variations significantly impact only liquids with low viscosity, and their overall effect on relaxation time remains minor (less than 5%). Thus, although the effects of flow rate and OD are not highly significant, it remains advisable to select lower flow rates and smaller OD capillaries for samples with low viscosity, and vice versa. For PD measurements, variations in geometry and operating parameters - specifically, two capillary diameters and four different flow rates—reveal that while setup conditions do influence $\tau_r$, their effect is minor compared to the dominant role of fluid viscosity. Altering these parameters by up to an order of magnitude results in less than a 5% change in $\tau_r$, which is negligible relative to the viscosity-induced variations (see Table 2).

For the WP method, although the geometry of the plate and ring remains lim-ited across all measurements, the surface tension (SFT) is primarily influenced by immersion depth, immersion speed, and withdrawal speed [37]. To isolate the effect of viscosity, these parameters were meticulously controlled and held constant throughout all experiments. While variations in these setup conditions can affect absolute SFT values, their impact on the relaxation time ($\tau_r$) remains minor relative to the dominant influence of viscosity [33, 37]. Consistently maintaining these parameters minimized experimental variability and ensured that the observed trends in $\tau_r$ are attributed primarily to changes in fluid viscosity.

In the case of BP, as mentioned earlier, we utilized five different capillaries. However, the results obtained show only a marginal difference between the values of $m$ and $b$. It is observed that an increase in the capillary diameter results in a reduction of both $m$ and $b$ values. Consequently, we have chosen to present only the results obtained for the smallest and largest capillaries. The reduction of the capillary diameter from 0.433 mm (SH2040) to 0.205 mm (SH2031) leads to a decrease in both $m$ and $b$. The primary reason is that decreasing the diameter of the capillaries increases the pressure across the interface, which, in turn, reduces the relative contribution of additional pressure due to viscous forces of the liquid. For the BP method, despite more than twofold variation in capillary diameter, the influence on relaxation time ($\tau_r$) remains relatively minor. Experiments conducted with four different capillaries reveal that, when all other parameters are held constant, doubling the capillary diameter leads to only a 5–7% change in $\tau_r$. While capillary geometry does affect transient pressure profiles during bubble formation, the characteristic relaxation behavior remains largely unaffected, underscoring the dominant role of fluid properties over setup-induced effects.

As observed across all three methods - PD, WP, and BP - the transient measurements are influenced by setup parameters; however, the effect on the relaxation time $\tau_r$ remains limited, especially when varying only the dimensions of the measuring probe while keeping other operating parameters unchanged. Considering that the relaxation times for PD, BP, and WP methods scale roughly as $10^0$, $10^1$, and $10^2$ seconds respectively, the limited variation in $\tau_r$ due to setup changes makes it relatively independent of probe geometry within a reasonable range. Consequently, a relative comparison between these methods is sensible without detailed consideration of probe dimensions, provided that the probe sizes are not drastically different.

The following section compares different methods and provides a comprehensive understanding for choosing the most suitable method for measuring surface tension in liquids with specific viscosity.

*4.4. Phase plot analysis for optimal method selection*

All three techniques can measure equilibrium surface tension given sufficiently long surface age (e.g., 100 s for a liquid with 1000 mPa·s viscosity). For such a liquid, PD reaches equilibrium within 1 s, whereas BP and WP require minimum surface ages of 11 s and >80 s, respectively. While BP and WP can yield accurate equilibrium values when operated over extended timescales, rapid measurement at young surface ages is often critical for applications involving time-sensitive batch processing, volatile or high-viscosity liquids, and surfactant-containing systems such as those encountered in inkjet printing, pharmaceuticals, and food processing. Hence, comparing the minimum measurement time as a function of viscosity becomes essential for selecting the most suitable technique.

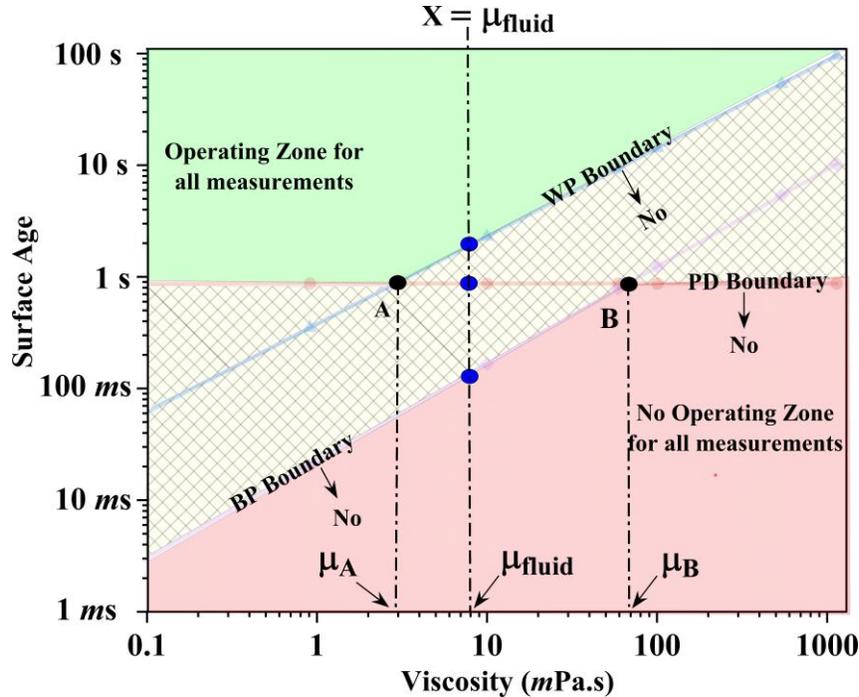

Figure 3: Phase plot: Variation of $\tau_r$ with viscosity ($\mu$) for pendant drop (15 $\mu$L/s), Wilhelmy plate and bubble pressure (smallest SH2031 capillary) methods. Pink region: no operating zone for all methods, green region: operating zone for all methods, and cross-hatch region: method dependent measurement zone.

Utilizing experimental results, Figure 3 can be illustrated as a phase plot encompassing data from three distinct methods. The plot serves the practical purpose of discerning the appropriate measurement technique for a given fluid viscosity, denoted as $\mu_{\text{fluid}}$, in conjunction with the surface age (same as $\tau_r$) of the DSFT measurement. It is important to emphasize that only data from the smallest capillaries of BP and Pendant drop at a flow rate of 15 $\mu L/s$ is employed for this thorough analysis. The intersections of these three boundaries identify six different zones. Each line on the plot delineates the measurement-method specific boundary between the operating and non-operating zones, labeled as BP Boundary, PD Boundary, and WP Boundary.

Standard operating protocol (SOP) for using the phase plot in figure 3 to determine accessible surface age for DSFT measurements

1. Step 1 – Identify the vertical reference line:
   (a) Draw a vertical line in the phase plot parallel to the Y-axis at $X = \mu_{fluid}$.
   (b) This line represents the specific fluid viscosity of interest ($\mu_{fluid}$) and is labeled as $X = \mu_{fluid}$ in Figure 3.

2. Step 2 – Locate intersections with method boundaries:
   (a) Along the $X = \mu_{fluid}$ line, identify three intersections corresponding to the boundaries of each measurement method:
      i. First intersection (border of the "No Operating Zone"):
         - Defines the minimum accessible surface age for accurate surface tension measurement, independent of the technique.
         - Example: For $\mu_{fluid} \approx 5$ mPa $\cdot$ s, the minimum time is approximately 70 ms, with BP being the first available method.
      ii. PD intersection:
         - For $\mu_{fluid} \approx 5$ mPa $\cdot$ s, PD becomes operable at 0.85 s.
      iii. WP intersection:
         - At the same viscosity, WP becomes operable at 1.1 s.
   (b) *Note:* If measurements must be performed with PD or WP instead of BP, use the corresponding intersection values as the minimum accessible surface ages for those methods.

3. Step 3 – Apply an engineering safety factor:
   (a) To account for experimental uncertainties, apply a 10% safety margin to the minimum accessible surface age values obtained from the phase plot.
   (b) Example: If the calculated minimum time is 0.85 s for PD, use $0.85 \times 1.10 \approx 0.94$ s as the operational minimum.
   (c) This ensures reliable performance under slight variations in experimental parameters.

4. Step 4 – Influence of flow rate, probe characteristics, and measurement–fit deviations:

(a) PD: Variations in flow rate or capillary diameter alter the relaxation time by less than 10%, a deviation fully accommodated by the 10% safety factor. This margin also compensates for minor deviations when measured data do not align exactly with the fitted curve.

(b) BP: Changes in capillary diameter affect the relaxation time by less than 10%, which is similarly encompassed by the 10% safety factor. This margin likewise corrects for small discrepancies between measured data and the fitted curve.

(c) WP: Variations in plate dimensions or material properties influence the relaxation time by less than 10%, also within the 10% safety factor. This margin also offsets minor deviations between measured data and the fitted curve.

5. Step 5 – Determine reliable operating regions:

    (a) Zones above each boundary line indicate the Operating Zone for that method, where reliable data can be obtained at the corresponding surface ages.

    (b) Example: For $\mu_{\text{fluid}} \approx 5 \text{ mPa} \cdot \text{s}$, all three methods (BP, PD, WP) perform reliably for measurement times > 2 s.

6. Step 6 – Recognize the three measurement regimes based on viscosity: In the zone between the "Operating Zone" and the "No Operating Zone" in the phase plot, the variation in required measurement time for each method can be categorized into three regimes:

    (a) Low-viscosity regime ($\mu < \mu_A \approx 3 \text{ mPa} \cdot \text{s}$):
    - BP provides the youngest accessible surface ages.
    - PD requires the longest time before measurement is possible.

    (b) Intermediate-viscosity regime ($\mu_A < \mu < \mu_B \approx 70 \text{ mPa} \cdot \text{s}$):
    - BP again offers the youngest possible surface ages.
    - WP requires the maximum relaxation time, setting the highest limit for accessible surface age.

    (c) High-viscosity regime ($\mu > \mu_B$):
    - PD outperforms BP and provides the youngest accessible surface ages.

7. Step 7 – Select the optimal method:

    (a) For a given $\mu_{\text{fluid}}$, choose the technique that lies above the operating boundary at the desired surface age.

    (b) Prioritize methods with shorter minimum accessible surface ages, ensuring the 10% safety factor is included for robust and reproducible measurements.

## 5. Conclusion

In conclusion, our study introduces a definitive parameter that quantifies the time required for reliable measurement of temporal surface tension variation with pendant drop, Wilhelmy plate and bubble pressure. The current study not only expands beyond previous studies of the drop volume method [58], but also provides a relative comparison. Despite the fact that viscosity does not affect surface tension, our research reveals that viscosity can lead to apparent deviations in surface tension measurements, depending on the method and measurement parameters. The results from this work is complementary to preliminary findings of pendant drop measurement method [28, 29, 30, 36, 60, 31] and bubble pressure tensiometer [22, 41]. Our key findings reveal empirical relationships between relaxation time and viscosity for surface tension measurements using three methods. These relationships show that equilibrium time varies linearly with viscosity on a logarithmic scale. For $\mu < 100$, mPa.s, no single method is preferable because the relaxation time $t_r$ is dominated by start-up effects. For $\mu > 100$ mPa.s, $t_r$ scales - $\sim 10^0$ (PD), $\sim 10^1$ (BP), $\sim 10^2$ (WP) - and this ordering persists under changes in setup geometry. The Wilhelmy plate method exhibits the steepest slope, indicating it requires the longest time for reliable results in practical liquids with viscosities greater than 3 mPa.s. Conversely, the bubble pressure method, with the smallest capillary, yields the shortest relaxation times for lower viscosity ranges of $0.1 \leq \mu \leq 70$ mPa.s. The pendant drop method performs best in higher viscosity regimes ($\mu > 70$ mPa.s). This relationship between relaxation time and viscosity is particularly critical when measuring the dynamic surface tension of highly viscous liquids with surface-active molecules or pure liquids, as viscosity can introduce inaccuracies in the measurements. The phase plot we provide can guide the selection of an appropriate method based on viscosity and the desired surface age to obtain reliable results where surface tension measurement is independent of liquid viscosity. As our findings suggest, the PD method performs best for high-viscosity liquids, making its extension to even higher viscosities a natural direction for further exploration. Future research should also focus on extending the current framework by integrating surfactant adsorption and diffusion effects alongside viscous relaxation. Superimposing these contributions will be essential for expanding the applicability of this comparative methodology from pure liquids to more practical multicomponent liquid samples, enabling a unified understanding of dynamic surface tension behavior. Additionally, consider incorporating alternative methods like drop volume [58, 61], spinning drop [62, 63], ring method [64, 65], etc.

## 6. Author contributions

**Vivek Kumar** - Data recording, analysis, formal analysis, writing - original draft, review, and editing. **JSM Quintero** - BP100 Data recording and analysis. **Aleksey Baldygin** - Data recording and analysis, writing - original draft, review, and editing. **Paul Molina** - BP100 Data recording and analysis. **Thomas Willers** - Conceptualization,, writing - original draft, review and editing. **Prashant Waghmare** - Conceptualization, funding acquisition, writing - original draft, review, and editing.

## 7. Conflicts of interest

The authors declare no conflict of interest.

## 8. Acknowledgments

The authors would also like to express gratitude to Adam and Ryan Baily. The authors thank **Natural Sciences and Engineering Research Council (NSERC)** for the financial support. "This research was supported by funding from the Canada First Research Excellence Fund as part of the University of Alberta's Future Energy Systems research initiative."

## 9. Data Availability

Raw data are available upon request to the Editor or the corresponding author.